\documentclass[aip,pop,10p,sanserif,reprint,twocolumn]{revtex4-1}
\usepackage[english]{babel}
\usepackage{amsmath, amsthm, amssymb, latexsym, graphicx}
\usepackage[utf8]{inputenc}
\usepackage{color} % Font Color 
\usepackage{enumerate}
\usepackage{lineno}
\newcommand{\del}{\delta}
\newcommand{\om}{\omega}

\begin{document}

\title{Two-fluid turbulence including electron inertia}

\author{Nahuel Andr\'es}
\email[E-mail: ]{nandres@iafe.uba.ar. Corresponding author: Av. Cantilo 2620 Edificio IAFE (CP 1428), CABA, Argentina. Tel.: +5411 47890179 (int. 134) ; Fax: +5411 47868114.}
\affiliation{Instituto de Astronom\'ia y F\'isica del Espacio, CC. 67, suc. 28, 1428, Buenos Aires, Argentina}
\affiliation{Departamento de F\'isica, Facultad de Ciencias Exactas y Naturales, Universidad de Buenos Aires, Pabell\'on I, 1428, Buenos Aires, Argentina}

\author{Carlos Gonzalez}
\affiliation{Departamento de Fisica, Facultad de Ciencias Exactas y Naturales, Universidad de Buenos Aires and IFIBA, CONICET, Buenos Aires, 1428, Argentina}

\author{Luis Martin}
\affiliation{Departamento de Fisica, Facultad de Ciencias Exactas y Naturales, Universidad de Buenos Aires and IFIBA, CONICET, Buenos Aires, 1428, Argentina}

\author{Pablo Dmitruk}
\affiliation{Departamento de Fisica, Facultad de Ciencias Exactas y Naturales, Universidad de Buenos Aires and IFIBA, CONICET, Buenos Aires, 1428, Argentina}

\author{Daniel G\'omez}
\affiliation{Instituto de Astronom\'ia y F\'isica del Espacio, CC. 67, suc. 28, 1428, Buenos Aires, Argentina}
\affiliation{Departamento de F\'isica, Facultad de Ciencias Exactas y Naturales, Universidad de Buenos Aires, Pabell\'on I, 1428, Buenos Aires, Argentina}

\date{\today}

\begin{abstract}
We present a full two-fluid magnetohydrodynamic (MHD) description for a completely ionized hydrogen plasma, retaining the effects of the Hall current, electron pressure and electron inertia. According to this description, each plasma species introduces a new spatial scale: the ion inertial length $\lambda_{i}$ and the electron inertial length $\lambda_{e}$, which are not present in the traditional MHD description. In the present paper, we seek for possible changes in the energy power spectrum in fully developed turbulent regimes, using numerical simulations of the two-fluid equations in two-and-a-half dimensions (2.5D). We have been able to reproduce different scaling laws in different spectral ranges, as it has been observed in the solar wind for the magnetic energy spectrum. At the smallest wavenumbers where plain MHD is valid, we obtain an inertial range following a Kolmogorov $k^{-5/3}$ law. For intermediate wavenumbers such that $\lambda_{i}^{-1} << k << \lambda_{e}^{-1}$, the spectrum is modified to a $k^{-7/3}$ power-law, as has also been obtained for Hall-MHD (HMHD) neglecting electron inertia terms. When electron inertia is retained, a new spectral region given by $k > \lambda_{e}^{-1}$ arises. The power spectrum for magnetic energy in this region is given by a $k^{-11/3}$ power law. Finally, when the terms of electron inertia are retained, we study the self-consistent electric field. Our results are discussed and compared with those obtained in solar wind observations and previous simulations.
\end{abstract}

\maketitle

\section{Introduction}\label{intro}

There are several alternative and complementary approaches to model the dynamics of a plasma. For instance, kinetic theory describes a plasma from a microscopic point of view, including phenomena at the corpuscular scales described through their distribution function. On the other hand, MHD models describe more global phenomena at macroscopic scales using low order moments of the distribution function such as the particle density, the velocity of the flow and its pressure. However, the description at intermediate scales, i.e. in between the MHD and the kinetic scales are a subject of debate. In particular, the solar wind is an example of a space plasma for which this discussion is still open. With the aim of getting a better understanding of the nature of a magnetized plasma at intermediate scales and within the framework of a full two-fluid MHD description, retaining the effects of the Hall current, electron pressure and electron inertia, we investigate the development of turbulent regimes throughout these 
scales.

An important feature to characterize a stationary and isotropic turbulent regime of a plasma is its energy spectrum $E(k)$, which provides the energy per unit wavenumber. At MHD scales, i.e. at wavenumber below the inverse of the ion inertial length $k_{\lambda_i}\sim\lambda_i^{-1}$ ($\lambda_i\equiv c/\omega_{pi}$ where $c$ is the speed of light and $\om_{pi}=(4\pi e^2n_0/m_i)^{1/2}$ is the ion plasma frequency), the energy spectrum follows a $k^{-5/3}$ scaling, i.e. a Kolmogorov spectrum just as for neutral fluids. This power-law was predicted by \citet{K1941} for hydrodynamic turbulence, assuming isotropy and using dimensional analysis. Using measurements of the solar wind at 1, 2.8 and 5 AU and assuming the Taylor hypothesis, \citet{M1982} found energy spectra consistent with a Kolmogorov spectrum. However, one fundamental difference between hydrodynamic turbulence and plasma turbulence is the presence of different wavenumber regimes with their corresponding power-law dependencies. Solar wind 
observations have shown that the MHD range typically ends just at the ion inertial length, where the magnetic power spectrum exhibits a characteristic break \citep{L2000,S2001}. At wavenumbers larger than the inverse of the ion inertial length, i.e the Hall-MHD range, the magnetic spectra exhibit steeper power laws \citep{G1994,G1996,L1998,S2006}. \citet{B1999} studied the electron magnetohydrodynamic (EMHD) turbulence in 2D and 3D systems. In the EMHD approximation, asymptotically valid at spatial scales much smaller than the ion inertial length, the ions are regarded as static (because of their much larger mass) and electrons are the only species to carry the electric current. In this regime, \citet{B1999} found that the energy spectrum follows a $k^{-7/3}$ power law. This prediction was later confirmed by other numerical simulations \citep{K2004,G2006,G2008}. The classical explanation for this turbulence regime is that it is associated with a cascade process involving dispersive waves, such as ion-
cyclotron and/or whistler modes. In the context of 3D compressible MHD with and without the Hall effect, \citet{D2006} analyzed the behavior of the magnetic and electric field fluctuations. The authors found that the turbulent magnetic field was almost unaffected by the presence of the Hall term in Ohm's law, while the electric field is modified at scales smaller than the ion skin depth and close to the dissipation range. Furthermore, reconnection zones are identified, and the relative importance of each term in Ohm's law was studied in real space. In this direction, \citet{Sm2004} examined the influence of the Hall effect and the level of turbulence on the magnetic reconnection rate in 2.5D compressible Hall MHD. Their results indicate that the reconnection rate is enhanced both by increasing the Hall parameter and the amplitude of the turbulence. The idea that MHD turbulence may play an important role in a magnetic reconnection setup was first proposed by \citet{M1986} by adding turbulent fluctuations on a 
two-dimensional sheet pinch configuration. It is also important to remark that several studies have shown that the magnetic reconnection rate might still depend on the value of the Hall parameter \citep{S2008,Bi1997,Sm2004,Mob2005,Moa2005} or on the level of turbulent fluctuations \citep{M1986,Sm2004}. 

Recently, \citet{S2009} found evidence of two breakpoints in the magnetic energy spectrum from solar wind observations obtained with the multi-spacecraft Cluster. These results show a break at a wavenumber presumably consistent with the inverse ion inertial length $k_{\lambda_i}$ (as previously observed by \citet{L2000} and \citet{S2001}) and a second break at a wavenumber correlated with the electron gyroradius $\rho_e=v_\perp/\omega_{ce}$ (where the perpendicular velocity is calculated with the thermal velocity and $\om_{ce}\equiv eB_0/m_ec$ is the electron cyclotron frequency). However, in those observations the electron gyroscale $\rho_e$ was very close to the electron inertial length $\lambda_e$  (because $\beta_e\sim 1$), and therefore it was not clear to what characteristic scale can be attributed this breakpoint. The authors confirmed the Kolmogorov spectrum at MHD scales, a second power law $k^{-7/3}$ at HMHD scales above the ion inertial scale and a steeper power law $k^{-4.1}$ above $k_{\rho_e}\equiv1/\rho_e$. Other authors have found similar results \citep{A2009,S2010}. In particular, \citet{A2009} confirmed the Kolmogorov law at the MHD scales, a power law $k^{-2.8}$ at $k>k_{\lambda_i}$ and an exponential decay around the electron Larmor radius $\rho_e$.

In summary, the present study is consistent with results suggesting that the Hall effect produces a steepening in the spectrum at the ion inertial length which does not involve energy dissipation \citep{G1996,St2001,G2006,G2007}. The main goal of the present paper is to explore the physics of a complete two-fluid model without neglecting the electron mass. Consequently, the system is able to distinguish two characteristic scales, the ion and electron inertial lengths. Therefore, we call Electron Inertia Hall-MHD (EIHMHD) to a theoretical framework that extends MHD and HMHD, both of which can be regarded as particular cases. In particular, this level of description should not be confused with the EMHD approximation \citep{B1992,Bi1997}, since we retain the whole dynamics of both the electron and ion flows throughout all the relevant spatial scales. We claim that the EIHMHD framework is a way to partially bridge the gap between the fluidistic and kinetic descriptions.

The paper is organized as follows, in section \ref{eih-mhd} we develop the EIHMHD model and present the ideal invariants of the model in section \ref{invariants}. In section \ref{2.5D} we show the set of equations that describe the dynamical evolution of the problem in a 2.5D setup. In section \ref{results} we present our main results and, finally, in section \ref{conclus} we summarize our conclusions.

\section{Electron Inertia Hall-MHD model}\label{eih-mhd}

The equations of motion for an incompressible plasma made of ions and electrons with mass $m_{i,e}$, charge $\pm e$, density $n_{i}=n_{e}=n$ (quasi-neutrality), pressure $p_{i,e}$ 
and velocity $\textbf{u}_{i,e}$ respectively, can be written as

\begin{eqnarray}\label{motionp}
m_in\frac{d \textbf{u}_i}{dt} &=&  en(\textbf{E}+\frac{1}{c}\textbf{u}_i\times\textbf{B})-\boldsymbol\nabla p_i + \nonumber
\\ &+& \mu_i \nabla^2\textbf{u}_i+ R_{ie}
\end{eqnarray}

\begin{eqnarray}\label{motione}
m_en\frac{d \textbf{u}_e}{dt} &=&  -en(\textbf{E}+\frac{1}{c}\textbf{u}_e\times\textbf{B})-\boldsymbol\nabla p_e + \nonumber
\\ &+& \mu_e \nabla^2\textbf{u}_e+ R_{ei}
\end{eqnarray}

\begin{eqnarray}\label{ampere}
\textbf{j}&=&\frac{c}{4\pi}\boldsymbol\nabla\times\textbf{B}={en}(\textbf{u}_i-\textbf{u}_e)
\end{eqnarray}

Here $\textbf{B}$ and $\textbf{E}$ are the magnetic and electric fields, $\textbf{j}$ is the electric current density,  $c$ is the speed of light, $\mu_{i,e}$ are the viscosities and $R_{ie}$ ($R_{ie} = - R_{ei}$) is the rate of momentum gained by ions due to collisions with electrons. This momentum exchange rate is assumed to be proportional to the relative 
speed between species. More specifically,
\begin{equation}\label{Rie}
 R_{ie} = - nm_i\nu_{ie}(\textbf{u}_i - \textbf{u}_e)
\end{equation}
where $\nu_{ie}$ is the collisional frequency of an ion against electrons. In view of equation \eqref{ampere}, this momentum exchange rate (or friction force between species) becomes proportional to the electric current density $\textbf{j}$.

The total derivatives in Equations~(\ref{motionp})-(\ref{motione}) are
\begin{equation}
 \frac{d\textbf{u}_{i,e}}{dt} \equiv \frac{\partial\textbf{u}_{i,e}}{\partial t} + (\textbf{u}_{i,e}\cdot\boldsymbol\nabla)\textbf{u}_{i,e}
\end{equation}
and the conservation of mass for each species leads, in the incompressible case, to
\begin{equation}\label{mass}
 \boldsymbol\nabla\cdot\textbf{u}_{i,e} = 0
\end{equation}
This set of equations can be written in a dimensionless form in terms of a typical length scale $L_0$, the constant particle density $n$, a value $B_0$ for the magnetic field, a typical velocity $v_A=B_0/(4\pi nM)^{1/2}$ (the Alfv\'en velocity) where $M\equiv m_i+m_e$ and the electric field is in units of $E_0 = v_AB_0/c$,
\begin{eqnarray}\label{dlessp}
 (1-\del)\frac{d \textbf{u}_i}{dt} &=&  \frac{1}{\lambda}(\textbf{E}+\textbf{u}_i\times\textbf{B})-\boldsymbol\nabla p_i+\nu_i\nabla^2\textbf{u}_i+\frac{\textbf{r}}{\lambda} \\ 
\label{dlesse}
 \del\frac{d \textbf{u}_e}{dt} &=&  -\frac{1}{\lambda}(\textbf{E}+\textbf{u}_e\times\textbf{B})-\boldsymbol\nabla p_e +\nu_e\nabla^2\textbf{u}_e-\frac{\textbf{r}}{\lambda} \\ 
\label{dlesso}
\textbf{j} &=& \frac{1}{\lambda} (\textbf{u}_i-\textbf{u}_e) 
\end{eqnarray} 
where we have introduced the dimensionless parameters $\del\equiv m_e/M$ and $\lambda\equiv c/\om_{M}L_0$, and $\om_{M}=(4\pi e^2n/M)^{1/2}$  has the form of a plasma frequency for a particle of mass $M$. The dimensionless momentum exchange rate is $\textbf{r} = -\eta\textbf{j}$ and $\eta = m_ic^2\nu_{ie}/(4\pi e^2nv_A L_0)$ is the (dimensionless) electric resistivity. The dimensionless ion and electron inertial lengths can be defined in terms of their corresponding plasma frequencies $\om_{i,e}=(4\pi e^2n/m_{i,e})^{1/2}$ simply 
as $\lambda_{i,e}\equiv c/\om_{i,e}L_0$. Note that in the limit of electron inertia equal to zero, we obtain $\om_{M}=\om_{i}$, and therefore $\lambda = \lambda_i = c/\om_{i}L_0$ reduces to the usual Hall parameter. However, throughout this paper we are going to retain the effect of electron inertia through the parameter $\del \ne 0$. For a fully ionized hydrogen plasma is $\del\ll 1$ and as a result $\lambda \ne \lambda_i \gg \lambda_e$. Nonetheless, the current theoretical description may also be applied to an electron-positron plasma (for which $\del = 1/2$), since it is actually valid for all values of $\del$. The expressions for the dimensionless ion and electron inertial scales ($\lambda_{i,e}$) in terms of the two dimensionless parameters $\delta$ and $\lambda$ are simply $\lambda_i=(1-\del)^{1/2}\lambda$ and $\lambda_e=\del^{1/2}\lambda$.

For a hydrodynamic description of this two-fluid plasma, we replace the velocity field for each species (i.e. $\textbf{u}_{i,e}$) in terms of two new vector fields. Namely, the hydrodynamic velocity $\textbf{u}$ given by
\begin{equation}\label{hdvelocity}
 \textbf{u} = (1-\del)\textbf{u}_i+\del\textbf{u}_e
\end{equation}
and the electric current density $\textbf{j}$ given by \eqref{dlesso}. From equations \eqref{dlesso}-\eqref{hdvelocity}, we can readily obtain the velocity of each species as
\begin{eqnarray}
 \textbf{u}_i&=&\textbf{u} + \del\lambda\textbf{j} \\
 \textbf{u}_e&=&\textbf{u} - (1-\del)\lambda\textbf{j}
\end{eqnarray}
The hydrodynamic equation of motion is the sum of the corresponding equations of motion \eqref{dlessp} and \eqref{dlesse},
\begin{equation}\label{hydro}
\frac{d \textbf{u}}{dt} = \textbf{j}\times\left[\textbf{B}-\del(1-\del)\lambda^2\nabla^2\textbf{B}\right] - \boldsymbol\nabla p + \nu\nabla^2\textbf{u} + \nu_0\nabla^2\textbf{j}
\end{equation}
where $p\equiv p_i+p_e$ is the total pressure, $\nu = \nu_i + \nu_e$ 
and $\nu_0 = \lambda(\del\nu_i -(1-\del)\nu_e)$. Following the expressions obtained by \citet{B1965} and assuming both species to share a common temperature, the ratio of viscosities is only a function of the mass ratio, i.e.
\begin{equation}
 \frac{\nu_e}{\nu_i} = 0.54 \sqrt{\frac{\del}{1-\del}}
\end{equation}
which shows that viscosity is predominantly due to ions. 

Note that most of the terms in equation \eqref{hydro} can easily be identified as a sum of the corresponding terms in equations \eqref{dlessp}-\eqref{dlesse}, but the convective 
derivatives in these equations are nonlinear terms that have also been properly taken into account, giving rise to a new nonlinear term in equation \eqref{hydro} which is proportional to $\del$. Note also that in the limit of negligible electron inertia (i.e., for $\delta\rightarrow 0$), equation \eqref{hydro} reduces to the equation of motion for traditional MHD. This is the case as well for the Hall-MHD description, which is a two-fluid theoretical description, but considering massless electrons. 

On the other hand, the equation of motion for electrons (\ref{dlesse}), using $\textbf{E}=-\partial_t\textbf{A}-\nabla\phi$ and $(\textbf{u}_e\cdot\nabla)\textbf{u}_e=\boldsymbol\omega_e\times\textbf{u}_e+\nabla(u_e^2/2)$ (with $\boldsymbol\omega_e=\nabla\times\textbf{u}_e$ being the electron vorticity) can be written as 
\begin{eqnarray}\label{mote}
\frac{\partial}{\partial t}(\textbf{A}-\delta\lambda\textbf{u}_e) &=& \textbf{u}_e\times(\textbf{B}- \delta\lambda\boldsymbol\omega_e)+ \nabla(\lambda p_e+ \nonumber
\\ &+& \delta\lambda\frac{u_e^2}{2}-\phi) - \lambda\nu_e \nabla^2\textbf{u}_e - \eta\textbf{j}
\end{eqnarray}
We define,
\begin{eqnarray}\label{bp}
\textbf{B}'\equiv \textbf{B}-\delta\lambda\boldsymbol\omega_e&=&\textbf{B}-\delta (1-\delta) \lambda^2\nabla^2\textbf{B} -\delta\lambda\boldsymbol\omega
\end{eqnarray}
where $\boldsymbol\omega=\boldsymbol\nabla\times\textbf{u}$ 
is the hydrodynamic vorticity. Taking the curl of equation (\ref{mote}) it is possible to obtain a dynamical equation for the magnetic field 
\begin{eqnarray}\label{dynB}
\partial_t~\textbf{B}' &=& \boldsymbol\nabla\times{[\textbf{u}-(1-\delta)\lambda\textbf{j}]\times\textbf{B}'} + \eta\nabla^2\textbf{B} - \nonumber
\\ &-& \lambda\nu_e\nabla^2\boldsymbol\omega - (1-\delta)\lambda^2\nu_e\nabla^4\textbf{B}
\end{eqnarray} 
Equations \eqref{hydro} and \eqref{dynB} are the EIHMHD equations. It is interesting to note that the presence of the electron mass (and the corresponding viscosity coefficient $\nu_e$) introduces high order derivative terms that play the role of hyperviscosity. This certainly has an impact at large wavenumbers, affecting the distribution of energy at the small scales and the dissipative range of the energy power spectrum. The major source of dissipation of magnetic field in a plasma where the electron mass is not neglected, is the friction between the electrons themselves and not the loss of momentum of the electrons by collision with ions (as in the MHD and HMHD cases). This can be seen in the last term of equation (\ref{dynB}), which together with the second term (on the right hand side) came from the curl of the dissipative term in the fluid equation of electrons, a term that can not be neglected if we consider electron inertia (and the resulting momentum and energy transport due to the electrons). 

It is also possible to obtain an equation for the electric field \textbf{E} making use of equations \eqref{hydro}, \eqref{bp}, \eqref{dynB} and the Maxwell–Faraday equation (in dimensionless form),
\begin{equation}\label{faraday}
\boldsymbol\nabla\times \textbf{E} = - \frac{\partial \textbf{B}}{\partial t} 
\end{equation}
It is useful to consider this equation in Fourier space to obtain a closed expression for the electric field. First the partial time derivative of equation \eqref{bp} in Fourier space reads
\begin{eqnarray}
\frac{\partial\hat{\textbf{B}}'}{\partial t}=\alpha_k\frac{\partial \hat{\textbf{B}}}{\partial t}-\delta\lambda\frac{\partial \hat{\boldsymbol\omega}}{\partial t}
\end{eqnarray}
where $\alpha_k\equiv 1+(1-\delta)\delta\lambda^2k^2$ since $\boldsymbol\nabla\rightarrow i\textbf{k}$ and $\hat{\textbf{A}}(\textbf{k})$ is the Fourier transform of an integrable function $\bf{A}(\textbf{x})$. Rearranging terms and using equation \eqref{faraday} in Fourier space we get an equation for the electric field in Fourier space as
\begin{eqnarray}\label{efield}
 i\textbf{k}\times\hat{\textbf{E}}&=& \alpha_k^{-1}\bigg(\frac{\partial \hat{\textbf{B}}'}{\partial t}+\delta\lambda\frac{\partial \hat{\boldsymbol\omega}}{\partial t}\bigg)
\end{eqnarray}
where the two right-hand side terms are calculated from equations \eqref{hydro} and \eqref{dynB} respectively as,
\begin{eqnarray}\nonumber
 \frac{\partial \hat{\boldsymbol\omega}}{\partial t} &=& i\textbf{k}\times\widehat{(\textbf{j}\times\textbf{B}')} + i\delta\lambda\textbf{k}\times\widehat{(\textbf{u}_i\times\boldsymbol\omega)} \\ 
 &&-k^2(\nu\hat{\textbf{u}}+\nu_0\hat{\textbf{j}}) \\ \nonumber
 \frac{\partial \hat{\textbf{B}}'}{\partial t} &=& i\textbf{k}\times\widehat{(\textbf{u}_e\times\textbf{B}')} \\ 
 &&-k^2(\eta\hat{\textbf{B}}-\lambda\nu_e\hat{\boldsymbol\omega}) - (1-\delta)\lambda^2\nu_ek^4\hat{\textbf{B}} 
\end{eqnarray}
The equation for the electric field is obtained applying $(\boldsymbol\nabla\times)^{-1}$ to equation \eqref{efield}, which gives rise to the gradient of an undetermined function $g({\bf r},t)$, which can be associated to the electrostatic potential. This function $g({\bf r},t)$ can be obtained from the Poisson equation that results from taking the divergence of the equation.

It is worth mentioning that the electric field consists of four different contributions. An inductive part related to the $\textbf{u}\times\textbf{B}$ term, a Hall contribution 
related to $\lambda_i\textbf{j}\times\textbf{B}$ term, the dissipative component and a new contribution associated with the non-zero electron mass (i.e., proportional to $\delta$). 

\section{Ideal invariants and energy cascade regions}\label{invariants}

In the ideal limit, i.e. neglecting dissipation terms (see also \citet{A2014}), a multi-species plasma made of N species of individual mass $m_s$, electric charge $q_s$ and particle density $n_s$, satisfies the following equations of motion 
\begin{equation}\label{motions}
m_sn_s\frac{d \textbf{u}_s}{dt} =  q_sn_s(\textbf{E}+\frac{1}{c}\textbf{u}_s\times\textbf{B})-\boldsymbol\nabla p_s 
\end{equation}
where $s = 1,\dots , N$. We assume each species to be incompressible (i.e. $n_s = const$ and $\boldsymbol\nabla\cdot\textbf{u}_s = 0$, for $s=1,\dots,N$) and the plasma to be quasi-neutral, i.e.
\begin{equation}\label{quasi}
 \sum_{s=1}^N q_s n_s = 0
\end{equation}
The electric current density will be given by
\begin{equation}\label{j}
 \textbf{j} = \frac{c}{4\pi}\boldsymbol\nabla\times\textbf{B} = \sum_{s=1}^N q_s n_s \textbf{u}_s
\end{equation}
Such a plasma displays $N+1$ ideal invariants. One of them is of course the total energy $E$, given by
\begin{equation}\label{general_ene}
 E = \int d^3r \bigg(\sum_{s=1}^N\frac{m_sn_su^2_s}{2}+\frac{B^2}{8\pi}\bigg)
\end{equation}
The other invariants are one helicity per species, i.e. 
\begin{equation}\label{general_hel}
H_s = \int d^3r \left(\textbf{A}+\frac{cm_s}{q_s}\textbf{u}_s\right)\cdot\left(\textbf{B}+\frac{cm_s}{q_s}\boldsymbol\omega_s\right)
\end{equation}
where $\boldsymbol\omega_s=\boldsymbol\nabla\times\textbf{u}_s$. In a fully ionized hydrogen plasma is $s = i,e$ and we therefore have three ideal invariants. In the Hall-MHD limit we neglect the electron mass ($m_ e = 0$) and as a result the total energy reduces to just the ion (or bulk) kinetic energy plus the magnetic energy (see equation \eqref{general_ene}). Also, the electron helicity (see equation \eqref{general_hel} for $m_s = 0$) reduces to the well known magnetic helicity $H_0 = \int d^3r \textbf{A}\cdot\textbf{B}$, while the proton helicity corresponds to the hybrid helicity (see for instance \citet{T1986}, also \citet{G2008}).

Note that when the effects of electron inertia are retained (i.e. $m_e \ne 0$), the regular magnetic helicity $H_0$ is not anymore an ideal invariant. For the two-fluid description being addressed in the present study, the dimensionless version of the three ideal invariants is
\begin{eqnarray}
E &=& \int d^3r \big(\frac{u}{2}^2+\frac{B}{2}^2+(1-\del)\del\lambda^2\frac{j}{2}^2\big) \\
H_e &=& \int d^3r \big( \textbf{A}-\del\lambda\textbf{u}\big) \cdot \big( \textbf{B}-\del\lambda\boldmath{\omega}\big) \\
H_i &=& \int d^3r \big( \textbf{A}+(1-\del)\lambda\textbf{u}\big) \cdot \big( \textbf{B}+(1-\del)\lambda\boldmath{\omega}\big)
\end{eqnarray}
All these are quadratic and global invariants. For instance, the energy density
\begin{equation}\label{enedens}
 E (\textbf{r}, t) = \frac{u}{2}^2+\frac{B}{2}^2+(1-\del)\del\lambda^2\frac{j}{2}^2\ ,
\end{equation}
satisfies the following evolution equation
\begin{equation}\label{enedot}
 \frac{\partial}{\partial t} E (\textbf{r}, t) = -\boldsymbol\nabla\cdot\textbf{F}\ , 
\end{equation}
where $\textbf{F}$ is therefore the energy flux. Since the energy density \eqref{enedens} is quadratic, an equation equivalent to \eqref{enedot} also holds in Fourier space as a result of Parseval's theorem. In a stationary and isotropic turbulent regime, the so called energy cascade corresponds to a constant energy flux in Fourier space (i.e. $\textbf{F}_k$ independent of $k=|\textbf{k}|$), which is therefore equal to the energy dissipation rate $\epsilon$. For instance, in the paradigmatic case of incompressible hydrodynamic turbulence, the modulus of the energy flux in Fourier space goes like $F_k \simeq k u_k^3 = \epsilon$, which leads to the well known 
Kolmogorov's energy power spectrum $E_k \simeq \epsilon^{2/3}k^{-5/3}$, simply using that $E_k\simeq u_k^2/\tau_k$ and $\tau_k\simeq (ku_k)^{-1}$. 

In the more complex case of EIHMHD turbulence, there are many terms contributing to the energy flux in both physical and Fourier spaces. Symbolically, these various contributions are sketched in the following expression for the energy flux in Fourier space,
\begin{equation}
F_k\simeq k(u_k^3+u_kB_kB'_k+(1-\delta)\lambda J_kB_kB'_k+(1-\delta)\delta\lambda^2\partial_tJ_kB_k) 
\end{equation}
The presence in EIHMHD of two physical lengthscales causes the appearance of three different regions in wavenumber space.
\begin{enumerate}[I]
\item MHD region ($k \lesssim k_{\lambda_i}$): In this region we assume $\del \approx 0 \approx \lambda$ and also $u_k\simeq B_k\simeq B'_k$. Therefore $F_k\simeq  kB_k^3=\epsilon$ and $E_k\simeq B_k^2/k \simeq \epsilon^{2/3}k^{-5/3}$.
\item HMHD region ($k_{\lambda_i} \lesssim k \lesssim k_{\lambda_e}$): In this region we maintain $\del\approx 0$ but $\lambda\ne 0$, and $u_k\lesssim B_k \simeq B'_k$. As a result, we now have $F_k\simeq \lambda k^2B_k^3=\epsilon$ and therefore $E_k\simeq B_k^2/k\simeq (\epsilon/\lambda)^{2/3} k^{-7/3}$.
\item EIHMHD region ($ k_{\lambda_e} \lesssim k$): This large-k region is dominated by the last two terms in the energy flux, i.e. $F_k\simeq k\del\lambda^2\partial_tJ_kB_k=\epsilon$ (since $B_k'\sim\del\lambda^{2}k^{2}B_k>>B_k$) . Since $k\lambda\gtrsim1/\sqrt{\delta}>>1$ we assume the ions to remain static because of their much larger mass and the dynamics to be dominated by the electrons, i.e.  $\partial_t \simeq k u_{ek} \simeq \lambda k^2 B_k$. Therefore $F_k\simeq\del\lambda^3k^4B_k^3=\epsilon$. Note that the energy power spectrum in this region is now predominantly electron kinetic energy, and therefore $E_k\simeq \del\lambda^2J_k^2/k\simeq (\epsilon^2\del)^{1/3} k^{-5/3}$. The power spectrum of magnetic energy, however, is equal to $B_k^2/k\simeq (\epsilon/(\del\lambda^3))^{2/3} k^{-11/3}$.
\end{enumerate}

\section{2.5D Setup}\label{2.5D}

We consider a 2.5D setup where the vector fields depend on two Cartesian coordinates, say \textit{x} and \textit{y}, but they have all three components. Considering the incompressible case, i.e. $\boldmath\nabla\cdot\textbf{u}=0$, we can write the magnetic and velocity fields as $\textbf{B}= \boldsymbol\nabla\times[\hat{\textbf{z}} a(x,y,t)] + \hat{\textbf{z}}b(x,y,t)$ and $\textbf{u}=\boldsymbol\nabla\times[\hat{\textbf{z}}\varphi(x,y,t)] + \hat{\textbf{z}}u(x,y,t)$, where $a(x,y,t)$ and $\varphi(x,y,t)$ are the scalar potential for the magnetic and velocity fields respectively. In terms of these scalar potentials, the equations (\ref{hydro}) 
and (\ref{dynB}) take the form
\begin{eqnarray}
\label{1.1}
\partial_t~\omega &=& [\varphi,\omega] - [a,j] - (1-\del)\del\lambda^2[b,\nabla^2b] + \nonumber
\\ &+& \nu\nabla^2\omega - \nu_0\lambda\nabla^4 b \\
\label{1.2}
\partial_t~u &=& [\varphi,u] - [a,b] - (1-\del)\del\lambda^2[j,b] + \nu\nabla^2 u + \nonumber
\\ &+& \nu_0\lambda\nabla^2 j \\
\label{1.3}
\partial_t~a' &=& [\varphi_e,a']+ \eta\nabla^2a - (1-\del)\nu_e\lambda^2\nabla^4 a - \nonumber
\\ &-& \nu_e\lambda\nabla^2 u \\
\label{1.4}
\partial_t~b' &=& [\varphi_e,b'] + [u_e,a'] + \eta\nabla^2 b - (1-\del)\nu_e\lambda^2\nabla^4 b - \nonumber
\\ &-& \nu_e\lambda\nabla^2 \omega 
\end{eqnarray}
where $\omega=-\nabla^2\varphi$, $j=-\nabla^2a$, $a'=a+\del(1-\del)\lambda^2j-\del\lambda u$, $b'=b-\del(1-\del)\lambda^2\nabla^2b-\del\lambda\omega$, 
and the nonlinear terms are the standard Poisson brackets, i.e. $[p,q]=\partial_xp\partial_yq-\partial_yp\partial_xq$. We have also defined the stream function and the velocity component along $\hat{\textbf{z}}$ for electrons, respectively as $\varphi_e = \varphi - (1-\del)\lambda b$ and $u_e = u - (1-\del)\lambda j$. This set of equations describes the dynamical evolution of the magnetic and velocity fields. When $\delta=0$ (i.e. $m_e=0$) it reduces to the incompressible 2.5D HMHD equations. Finally, in the 2.5D setup, for computation of the self-consistent electric field along the $z$ direction we can ignore the $g({\bf r},t)$ indetermination since $\partial_z\equiv0$ in this geometry.

\section{Numerical results}\label{results}

We use a parallel pseudospectral code to numerically integrate equations \eqref{1.1}-\eqref{1.4}. A second-order Runge-Kutta time integration scheme is adopted. Periodic boundary conditions are assumed for the $\hat{\textbf{x}}$ and $\hat{\textbf{y}}$ directions of a square box of linear side $2\pi L_0$ (where $L_0$ is the length unit). The simulations performed throughout the present paper are run-down, i.e. they do not contain any magnetic or velocity  stirring forces. As initial conditions, we excite Fourier modes (for both magnetic and velocity field fluctuations) in a shell in $k$-space with wavenumbers $3 \le k \le 4$, with the same amplitude and random phases for all modes. For all the simulations presented here, we used a spatial resolution of $3072^2$ grid points, $\nu=3\times 10^{-5}$ and $\eta=1.5\times10^{-4}$. To suppress aliasing effects, our spectral code uses a maximum wavenumber $k_{max}=N/3=1024$.
\begin{figure}
\centering
\includegraphics[width=.5\textwidth]{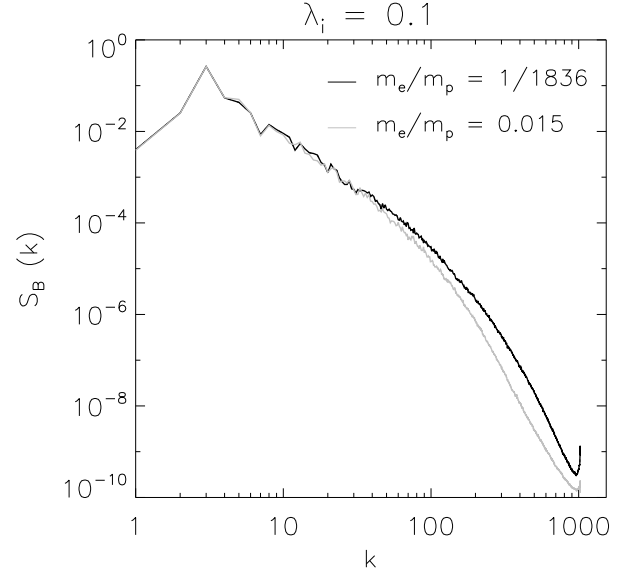}
\caption{Magnetic energy spectra for EIHMHD cases with $\lambda_i = 1/10$ and $m_e/m_p=1/1836$ (black) and $m_e/m_p=0.015$ (gray).}
\label{both}
\end{figure}
The ratio between the ion and electron inertial lengths are equal to the square root of the mass ratio. We used the realistic value $m_e/m_p=1/1836$ which correspond to $k_{\lambda_e}\sim43k_{\lambda_i}$. In addition, the dissipation range corresponds to wavenumbers much larger than these two characteristic scales. We ran simulations at high spatial resolution to study the freely evolving turbulence at different scales. In particular, we performed two EIHMHD simulations with the same ion inertial length ($\lambda_i$) and different electron to proton mass ratios ($m_e/m_p$). On one hand, we used a fictitious mass ratio, $m_e/m_p=0.015$ (electrons 27 times heavier), which corresponds to $k_{\lambda_i}\sim10$ and $k_{\lambda_e}\sim82$ to study the development of scales between the electron and the ion inertial lengths. On the other hand, we used the real mass ratio $m_e/m_p=1/1836$ corresponding to $k_{\lambda_p}\sim10$ and $k_{\lambda_e}\sim428$. For both simulations the dissipation wavenumber $k_d$, computed 
as $k_d=<j^2+\omega^2>^{1/4}/\sqrt{\nu}$, remains in the range of $k_e<k_d<k_{max}$. Figure \ref{both} shows the magnetic energy spectra for both cases. The black and gray lines correspond to the real and fictitious electron to proton mass ratio respectively. As shown by the spectra, the magnetic power spectra explicitly depends on the value of electron mass, even though asymptotically goes to the HMHD spectrum as $k<<k_{\lambda_e}$.
\begin{figure}
\centering
\includegraphics[width=.5\textwidth]{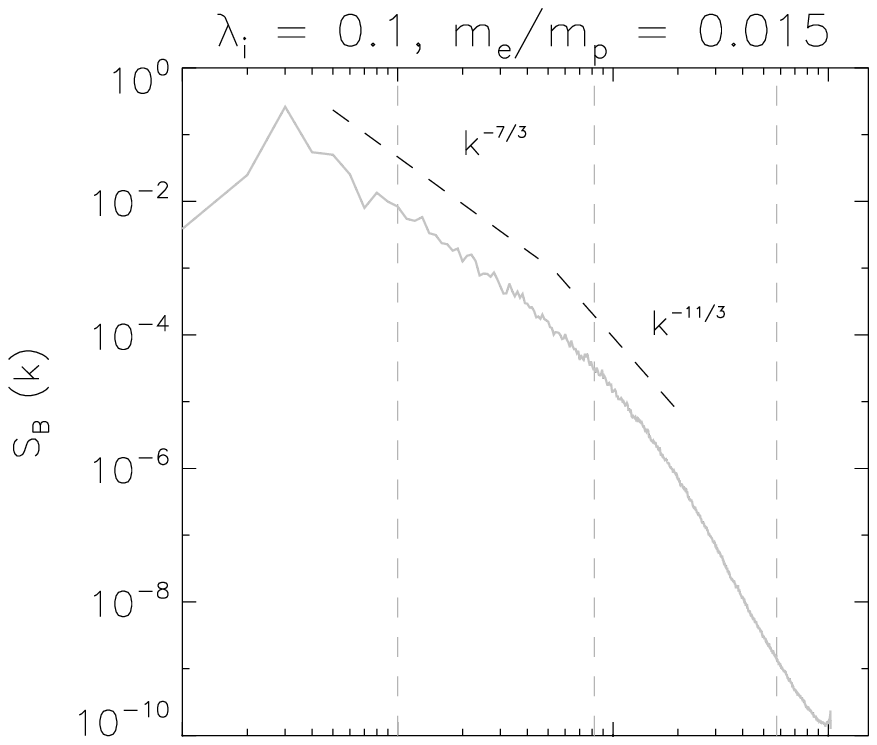}

\vspace{-1.2cm}
\includegraphics[width=.5\textwidth]{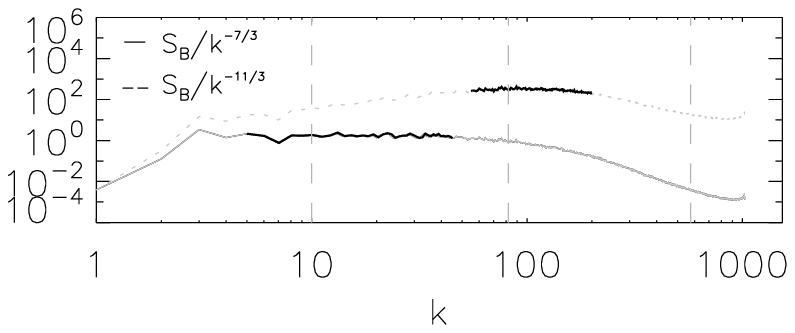}
\caption{Magnetic energy spectra for $m_e/m_p=0.015$. Vertical dashed gray lines correspond to $k_{\lambda_i}\sim10$, $k_{\lambda_e}\sim82$ and $k_{\nu}\sim550$. The compensated spectrum for the HMHD (solid line) and EIHMHD (dashed line) regions are shown in the lower panel.}
\label{eihmhd_f}
\end{figure}

The upper panel in Figure \ref{eihmhd_f} shows the magnetic energy spectrum for the case of fictitious electron to proton mass ratio (gray line). In addition, the dashed black lines show the theoretical power-law scalings (see section \ref{invariants}) for the different spectral ranges. The ion, electron and dissipation wavenumbers are indicated as vertical dashed gray lines. The lower panels show the compensated spectrum for the HMHD (solid line) and EIHMHD (dashed line) region. The separation points occur near the kinetic scales $k_{\lambda_i}$ and $k_{\lambda_e}$, which is consistent with solar wind observations \citep{S2009,A2009}. It is worth mentioning that both kinetic effects, the Hall effect and the non-zero electron mass, affect the spectrum and the breakpoints. It is also remarkable the consistent scaling for each region. From Figure \ref{eihmhd_f}, the scale separation between the HMHD and the EIHMHD regions is clearly noticeable. The Hall range is well described with the scaling $\sim k^{-7/3}$, 
in agreement with observations, several theoretical predictions \citep{G2006,B1999} and previous numerical results \citep{K2004,G2008}. A new range of scaling $\sim k^{-11/3}$ emerges for wavenumbers $k \sim k_{\lambda_{e}}$, i.e. the EIHMHD region, which is also consistent with our prediction, solar wind observations \citep{S2009,A2009} and previous simulations \citep{S2009,W2012}. There is also an indication of an exponential decay for the largest range of wavenumbers in our simulations, as was suggested by observations \citep{A2009}.
\begin{figure}
\centering
\includegraphics[width=.5\textwidth]{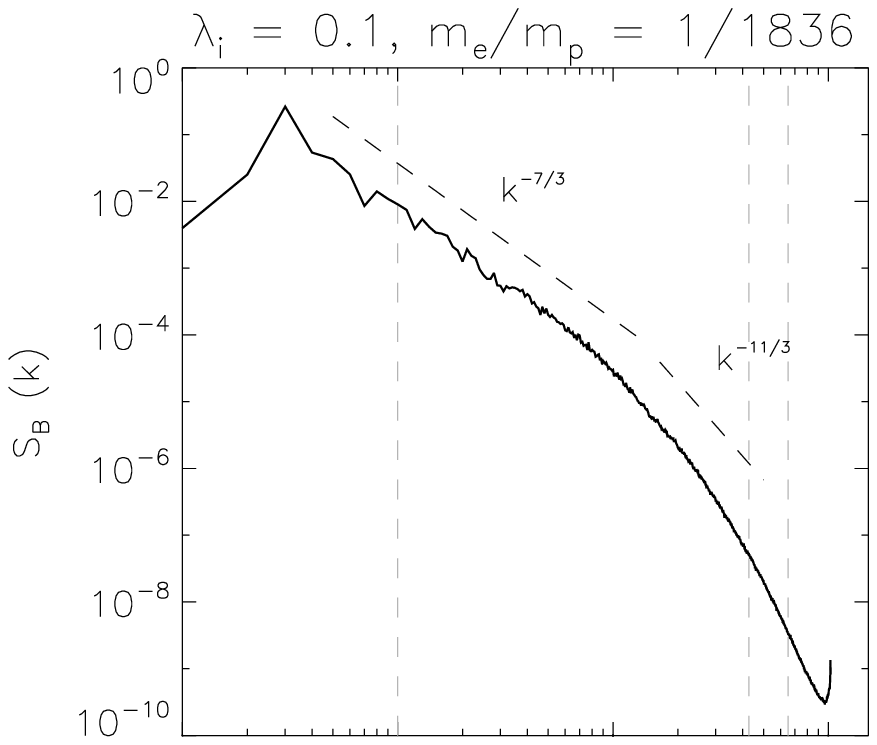}

\vspace{-1.2cm}
\includegraphics[width=.5\textwidth]{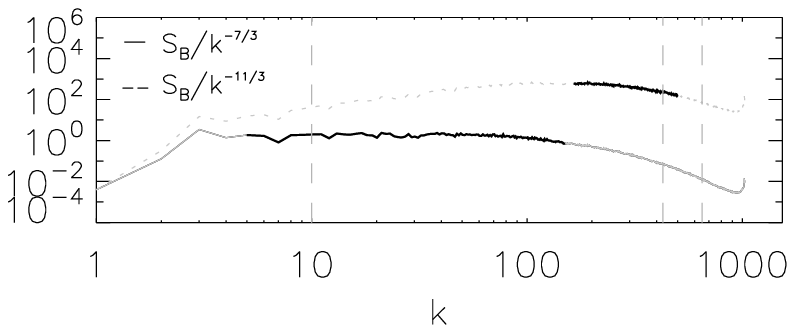}
\caption{Magnetic energy spectra for $m_e/m_p=1/1836$. Vertical dashed gray lines correspond to $k_{\lambda_i}\sim10$, $k_{\lambda_e}\sim430$ and $k_{\nu}\sim650$. The compensated spectrum for the HMHD (gray line) and EIHMHD (green line) regions in the same format as Figure \ref{eihmhd_f}.}
\label{eihmhd_r}
\end{figure}

Figure \ref{eihmhd_r} shows the power spectra for the magnetic energy for $m_e/m_p=1/1836$ (black line), with the same format as Figure \ref{eihmhd_f}. We also obtain two separation points at the two kinetic scales $k_{\lambda_i}$ and $k_{\lambda_e}$, with a $\sim k^{-7/3}$ and $\sim k^{-11/3}$ power-law scalings for HMHD and EIHMHD, respectively. However, the inverse of the electron inertial lengths and the dissipation wavenumber are close to each other in this case. Therefore, there is no clear-cut distinction between the $k^{-11/3}$  power-law and the exponentially decaying dissipative region.

\begin{figure}
\centering
\includegraphics[width=.5\textwidth]{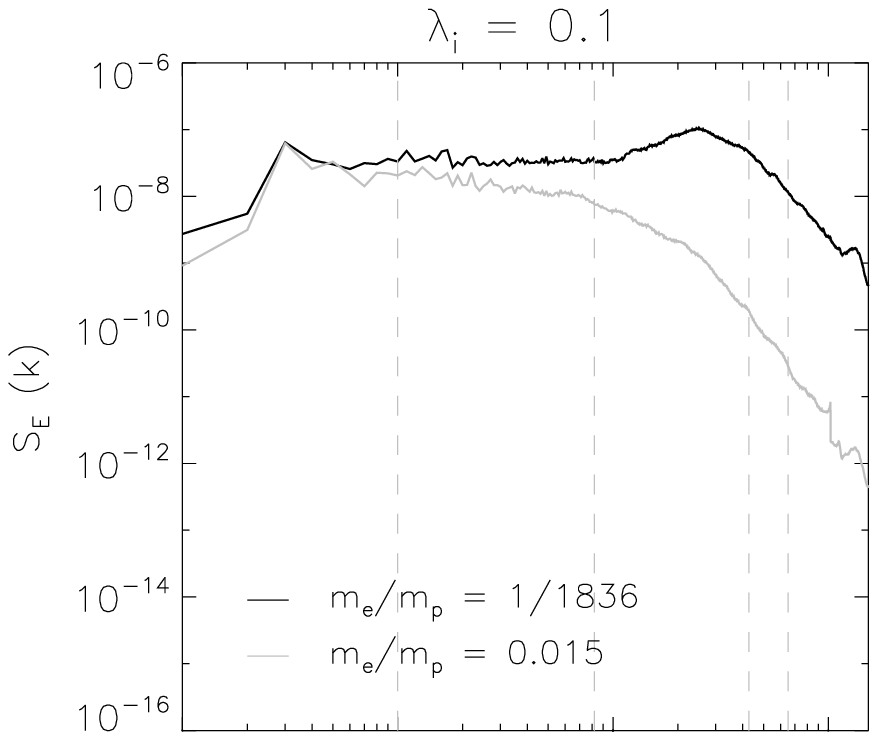}

\vspace{-1.2cm}
\includegraphics[width=.5\textwidth]{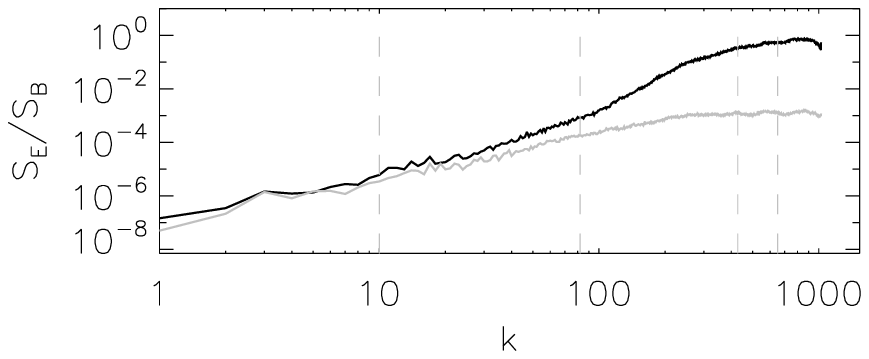}
\caption{Power spectrum of electrostatic field for EIHMHD with $m_e/m_p=1/1836$ and $m_e/m_p=0.015$ (upper panel). Vertical lines correspond to $k_{\lambda_i}\sim10$, $k_{\lambda_e}$ ($\sim82$ and $\sim428$ for the fictitious and real mass ratio respectively) and $k_{\nu}\sim650$. The lower panel, corresponds to the ratio between the electric and magnetic spectra, i.e. S$_\text{E}$/S$_\text{B}$.}
\label{ez}
\end{figure}
Figure \ref{ez} (upper panel) shows the power spectrum of the $z$ component of the electric field for the two EIHMHD cases,  $m_e/m_p=1836$ (black) and $m_e/m_p=0.015$ (gray). The ion wavenumber ($k_{\lambda_i}\sim10$), the fictitious and real electron wavenumbers ($k_{\lambda_e}\sim82$ and $k_{\lambda_e}\sim482$, respectively) and the dissipation scale ($k_d\sim650$) wavenumber are indicated as vertical dashed gray lines. The two spectra are clearly different when we consider electrons with different masses. As we expect, the electric field is much smaller than the magnetic field for all scales. The lower panel shows the ratio between the electric field ($z$ component) and the magnetic field spectra. We find that the electric field becomes gradually more important as $k$ increases. This is consistent with observations \citep{S2009} in the solar wind. 

\section{Conclusions}\label{conclus}

Within the context of a full two-fluid model we obtain different power laws for the magnetic energy spectrum, consistent with those observed in the solar wind. Allowing electrons to acquire a finite kinetic energy introduces a new range in the energy spectrum. According to our results, the separation points occur every time a new scale is involved, first the ion inertial length and then the electron inertial length. This is explicitly shown in equations (\ref{1.1})-(\ref{1.4}) where it can be seen that the presence of the scales $\lambda_{i}$ and $\lambda_{e}$ introduce new non-linear terms which are absent in a plain MHD description. As a consequence, these new nonlinear terms affect the energy distribution among scales. If the energy distribution is affected by introducing these two effects (Hall and non-zero electron mass), we can expect also different flow structures, intermittency and general dynamics on scales where we can not treat the plasma as a single fluid. We have taken a first step toward understanding turbulence in a 
full two-fluid model and leave the path for further studies of this system.

\section*{Acknowledgments}
We acknowledge support from the following grants PIP 11220090100825, UBACyT 20020110200359, 20020100100315, and PICT 2011-1529, 2011-1626, 2011-0454.

\end{document}